\documentclass[amssymb,superscriptaddress, twocolumn]{revtex4}
\usepackage{amsmath}
\usepackage{amssymb}
\usepackage{subfigure}
\usepackage{graphicx}
\usepackage{ulem}
\usepackage{multirow}
\usepackage{enumitem}
\usepackage{lipsum, babel}

\usepackage{amsmath}
\usepackage{amssymb}
\usepackage{subfigure}
\usepackage{graphicx}
\usepackage{ulem}
\usepackage{multirow}
\usepackage{enumitem}
\usepackage{lipsum, babel}
\newcommand{\be}{\begin{equation}}
	\newcommand{\ee}{\end{equation}}

\newcommand{\bea}{\begin{eqnarray}}
	\newcommand{\eea}{\end{eqnarray}}

\usepackage{xcolor}

\begin{document}

	\title{Topological gauge theory of vortices in type-III superconductors}% Force line breaks with \\
	
%	\author{M.\,C.\,Diamantini$^1$,\,C.\,A.\,Trugenberger$^2$,\, and V.\,M.\,Vinokur$^{3,\ast}$}
%	\author{Florian Neukart\,$^{1,2}$ \&
	%	Valerii Vinokur\,$^1$}%
	%\email{f.neukart@terraquantum.swiss}
	%\altaffiliation[Also at ]{Leiden Institute of Advanced Computer Science, Leiden University}%Lines break automatically or can be forced with \\
	% \author{Valerii Vinokur}%
	%\email{vv@terraquantum.swiss}
	%\affiliation{%
		%Terra Quantum AG\\
	
	% \collaboration{MUSO Collaboration}%\noaffiliation
	
	%  \homepage{http://www.Second.institution.edu/~Charlie.Author}
	% \affiliation{
		%  Second institution and/or address\\
		%  This line break forced% with \\
		% }%
	% \affiliation{
		%  Third institution, the second for Charlie Author
	% \affiliation{%
		%  Authors' institution and/or address\\
		%  This line break forced with \textbackslash\textbackslash
		% }%
	
	% \collaboration{CLEO Collaboration}%\noaffiliation
		\author{M.\,C.\,Diamantini}
	\affiliation{University of Perugia, via A. Pascoli, I-06100 Perugia, Italy}
	\author{C.\,A.\,Trugenberger}
	\affiliation{SwissScientific Technologies SA, rue du Rhone 59, CH-1204 Geneva, Switzerland}
	\author{V.\,M.\,Vinokur}
	\thanks{vv@terraquantum.swiss}
	\affiliation{Terra Quantum AG, Kornhausstrasse 25, CH-9000 St. Gallen, Switzerland}

	\date{\today}% It is always \today, today,
	
	%

	\begin{abstract}
			\noindent
		Traditional superconductors fall into two categories, type-I, expelling magnetic fields, and type-II, into which magnetic fields exceeding a lower critical field  $H_{\rm c1}$  penetrate in a form of Abrikosov vortices. Abrikosov vortices are characterized by two spatial scales, the size of the normal core, $\xi$, where superconducting order parameter is suppressed and the London penetration depth $\lambda$ describing the scale at which circulating superconducting currents forming vortices start to noticeably drop. Here we demonstrate that novel type-III superconductivity, realized in granular media in any dimension, %with a focus on its different 
		hosts a novel vortex physics. Type-III vortices have no cores, are logarithmically confined and carry only a gauge scale $\lambda$. Accordingly, in type-III superconductors $H_{\rm c1}=0$ at %$T=0$
		zero temperature and Ginzburg-Landau theory must be replaced by a topological gauge theory. Type-III superconductivity is destroyed not by Cooper pair breaking but by vortex proliferation generalizing the Berezinskii-Kosterlitz-Thouless mechanism to any dimension. 
	\end{abstract}
	
	\maketitle
%	\upsubscripts
%	\thispagestyle{fancy}
%	\lfoot{\parbox{\textwidth}{ \vspace{0.3cm}
%			\rule{\textwidth}{0.2pt}
%			\hspace{-0.2cm} \textsf{\scalefont{0.80}
	\iffalse
				$^1$University of Perugia, via A. Pascoli, I-06100 Perugia, Italy; $^2$SwissScientific Technologies SA, rue du Rhone 59, CH-1204 Geneva, Switzerland; $^3$Terra Quantum AG, Kornhausstrasse 25, CH-9000 St. Gallen, Switzerland;
				$^*$correspondence to be sent to vv@terraquantum.swiss
	\fi
%			}
	%		\vspace{-0.2cm}
%			\begin{center}{\scalefont{0.87} \thepage}\end{center}}} \cfoot{}
	%\tableofcontents
	\iffalse
		 \begin{strip}  
		\text{{Keywords: Type-III superconductivity, vortices, type-III vortices, Ginzburg-Landau theory,
				Berezinskii-Kosterlitz-Thouless mechanism.}}
		%	\begin{flushright}
			%		Konfucius
			%	\end{flushright}
	\end{strip}  
	\fi
	%\section{\label{sec:level1}Introduction}
	\bigskip
	\section{Introduction}~~\\
	\noindent
Superconductivity, a coherent quantum state existing on a macroscopic scale, is one of most fascinating quantum phenomena, see e.g.,\,\cite{tinkham}. Its rich phenomenology is usually described by the Ginzburg–Landau (GL) theory in terms of the order parameter, representing the macroscopic wave function of the superconducting condensate, which is characterized by two length scales, the coherence length $\xi$ and the London penetration depth $\lambda$. The energies corresponding to these lengths describe the gaps for two excitations, the Higgs mode, related to amplitude oscillations of the order parameter and the photons, which become massive by absorbing the phase of the order parameter via the Anderson-Higgs mechanism. The Higgs gap coincides with the superconducting gap, see\,\cite{higgsrev} for a recent comprehensive review. Nonetheless, at low temperatures it is a well-defined, undamped mode since it is unaffected by Coulomb interactions\,\cite{varma} and its decay to single-particle excitations is suppressed by inverse powers of the temperature\,\cite{kogan}. 
The Higgs mode in superconductors, however, has remained elusive for quite some time. The reasons are twofold. First, the Higgs mode does not couple linearly to external electromagnetic probes and is thus difficult to detect by standard techniques. Second, the energy gap corresponding to this mode lies in the meV scale and, thus, one needs a THz light source to have the Higgs mode excited. Such THz lasers, however, have become available only in the last decade, and, indeed, with this technological advance, the Higgs mode of superconductors got clearly identified\,\cite{laser}.

The behavior of superconductors and their response to magnetic field was then understood as being determined by the relation of these two characteristic scales, $\xi$ and $\lambda$. When $\xi>\sqrt{2} \lambda$, the Higgs core exceeds the gauge core, and magnetic fields $H$ cannot penetrate the superconductor. At some critical value $H_{\rm c}=1/(2e\lambda\xi)$, where $e$ is the charge of the electron, the magnetic field destroys superconductivity (we use natural units with light velocity $c=1$, Planck constant $\hbar = 1$ and te vacuum dielectric permittivity $\varepsilon_0=1$). These superconductors with $\xi>\sqrt{2}\lambda$, not accepting the magnetic field, are called type I superconductors. If $\xi<\sqrt{2} \lambda$, the Higgs core is typically the smallest one and magnetic fields exceeding  the so-called lower critical field $H_{\rm c1}=(\xi/\sqrt{2}\lambda)\textrm{ln}(\lambda/\xi)H_{\rm c}$ penetrate such a material in a form of Abrikosov vortices, which are the solitons of the Ginzburg-Landau functional\,\cite{tinkham}. Abrikosov vortices do exist in the interval $H_{\rm c1}<H<H_{\rm c2}$, where the upper critical field, $H_{\rm c2}=\sqrt{2}(\lambda/\xi)H_{\rm c}$, marks the destruction of superconductivity. These superconductors with $\xi<\sqrt{2}\lambda$ are called type II superconductors.
	
The origin of the energy gap in electronic spectrum in superconductors was given in the famous Bardeen-Cooper-Scsrieffer (BCS) theory\,\cite{BCS1957}. 
The BCS theory 
related the formation of this gap to the attraction between electrons located near the Fermi level and having opposite momenta and spins. This attraction caused by electron-phonon interaction results in formation of the so-called Cooper pairs and creates the energy gap, $\Delta\simeq k_{\text B}T_{\text c}$ in electron spectrum, where $k_{\text B}$ is Boltzmann constant and $T_{\text c}$ is the temperature of the superconducting transition. The energy necessary to break Cooper pairs according to the BCS theory is $E_{\text g}(T)=2\Delta(T)$, with the maximal value  $E_{\text g}(0)=2\Delta(0)=3.528k_{\text B}T_{\text c}$. This theoretical finding decisively agreed with the experiment.
	
	%The next step of our understanding of superconductivity was the establishment of the existence of the energy gap, $\Delta\simeq k_{\text B}T_{\text c}$, where $k_{\text B}$ is Boltzmann constant and $T_{\text c}$ is the temperature of the superconducting transition\,\cite{Tinkham,BCS1957}, and Bardeen, Cooper, and Schrieffer\,\cite{BCS1957} created a theory demonstrating that even a weak attractive interaction between electrons caused by the electron-phonon interaction results in instability in the ordinary Fermi ground state and in formation of the bound pairs of electrons, Cooper pairs, with equal and opposite momentum and spin. The prediction was that the minimum energy $E_{\mathtext g}(T)=2\Delta(T)$ is required to break a Cooper pair, and its limiting value is $E_{\mathtext g}(0)=2\Delta(0)=3.528k_{\text B}T_{\mathtext c}$. This decisively agreed with the experiment. 
	
	The breakthrough discovery of high-temperature superconductivity (HTS) in 1986 by Bednorz and Müller\,\cite{Bednorz} revealed a new kind of superconductor that is not explained by the BCS theory. The origin and underlying mechanism of HTS has remained a subject of heated debate for more than three decades after its discovery. The key to unraveling the nature of HTS
	came from resolving the enigma of the pseudogap state. The pseudogap state in
	the underdoped region is a distinct thermodynamic phase characterized by nematicity, temperature-quadratic resistive behavior and magnetoelectric
	effects. Till recently, a general description of the observed universal features of
	the pseudogap phase and their connection with HTS was lacking. A possible solution of the puzzle was recently proposed in\,\cite{DTV} where a topological field theory capturing all universal
	characteristics of the cuprate-based HTS materials and explaining the observed phase diagram was developed. The universality of the HTS phase diagram reflects a
	unique topological mechanism of competition between a dyon
	condensate and a Cooper pair condensate. 
	
%	The theory is based on the
%	fact that the unit cell of the cuprate-based HTS materials contains two neighboring CuO planes, which provides the pairing mechanism for Cooper pairs formation\,\cite{DTV-pairing}, making it possible for the Cooper
%	pairs not only to move along the ab-planes but also to tunnel between them in the c-direction. The theory explained two most amazing observations: (i) that even one unit cell-thin cuprate films have basically the same $T_{\text c}$ as bulk samples and (ii) that the distance between the Cooper pairs in the superconducting condensate well exceeds the size of the Cooper pair, $\simeq\xi$, where $\xi$ is the superconducting coherence length – opposite to what observed in conventional superconductors. 
	
	The HTS discovery suggested that the standard textbook type-I and type-II superconductors are not the only possibilities. And indeed, as was first shown in the mid 2010's\,\cite{topsc, higgsless}, there may exist one more type of superconductor. As we have demonstrated in\,\cite{typeIII}, this novel type-III superconductivity has only a gauge scale $\lambda$ and no Higgs core; thus, in these type-III superconductors, the lower critical field $H_{\rm c1}$ vanishes at $T=0$. In type\,III systems, the Cooper pair condensate, i.e., superconductivity, is destroyed at high temperatures not by the breaking up of Cooper pairs but 
	by the proliferation of vortices, which are logarithmically confined in the superconducting state. In 2D systems, this mechanism is immediately recognized as the Berezinskii-Kosterlitz-Thouless (BKT) mechanism of 2D superconductivity\,\cite{ber, kt}. This is indeed the prototype of type-III superconductors. With the appropriate modifications, this mechanism can take place in any dimension\,\cite{topsc, higgsless} and constitutes a genuine alternative type of superconductivity which has now been experimentally confirmed\,\cite{typeIII}. 
	\smallskip
	
	To understand the mechanism of type-III superconductivity, let us first discuss the 2D case, realized in thin films with thicknesses that are less or comparable with the coherence length $\xi$. Considering a superconducting film of thickness $d$, we have to remember two important differences with respect to the usual 3D case. First, in 2D, the electromagnetic coupling is $e^2$$_{\rm eff} = e^2/d$ and has canonical dimension [mass] (we  remind that we use natural units $c=1$, $\hbar = 1$, $\varepsilon_0=1$). Since it becomes ever stronger in the limit $d\to 0$, one cannot neglect the coupling of the Ginzburg-Landau theory to both magnetic and electric fluctuations. Secondly, the Anderson-Higgs gauge scale in thin films is the so-called Pearl length $\lambda_{\perp} = \lambda^2/d$, which exceeds typical system lateral sizes in the limit $d\to 0$. Therefore, the time-dependent Ginzburg-Landau model describing 2D superconductors is a super-renormalizable field theory with massless gauge fields (within the physical domain). Such field theories are plagued by infrared divergences \cite{jackiwIR}, which can be cured by a perturbative expansion in terms of ${\rm log} \left( e_{\rm eff}^2 L \right) =  {\rm log} \left( e^2 L/d \right) $, where $L$ is the system size, representing the necessary infrared cutoff. As a consequence, with decreasing $d$, a traditional superconductor will break up into islands of condensate having typical dimension $\ell = {\cal O}(d)$\,\cite{planar}. These islands have been detected already in the very first studies of superconducting films\,\cite{goldman1, hebard, goldmanrev} and are now a paradigm characterizing the structure of these systems\,\cite{sacepe1, sacepe2, granular}. These 2D systems are then typically modelled as Josephson junction arrays (JJAs)\,\cite{fazio, jjarev} with random couplings, see e.g.\,\cite{kapitulniknew}.

	\section{Consequences of the granular structure for vortices: quantum phase slips and tunnelling }~~\\

	%
	%Historically, the BCS theory, formulated by Bardeen, Cooper, and Schrieffer in 1957, provided the first microscopic explanation for superconductivity, postulating that electrons in certain metals could pair up, forming ``Cooper pairs", due to phonon-mediated attractions \cite{bardeen1957theory}. This theory reigned supreme until the surprising discovery of high-temperature superconductors in the 1980s, materials that exhibited superconductivity at temperatures much higher than previously thought possible \cite{bednorz1986possible}. This, along with the identification of other unconventional superconductors, presented new puzzles and inconsistencies with the traditional BCS theory.% 

		\vspace{-0.8cm}
			\noindent
	The self-induced electronic granularity immediately implies a fundamental difference of type-III superconductors from standard 3D homogeneous superconductors. Even if the individual grains are {\it local} superconductors, {\it global} superconductivity can be lost if the phase coherence and the ensuing tunnelling between grains are not realized, see\,\cite{book} for a review. The immediate question that arises is what are the relevant vortices in 
	granular superconductors. %this setting? %In principle, there could 
	The granular medium can host %be 
	two kinds of vortices, Abrikosov vortices within the individual superconducting islands, associated with local superconductivity, and what we call XY vortices, associated with global superconductivity. Notably, if the islands are small enough, i.e., have a size comparable with $\xi$ within them, the local, i.e., fitting within the granules, Abrikosov vortices are not stable\,\cite{likharev, gurevich}. 
	
\begin{figure}[t!]
		\includegraphics[width=7cm]{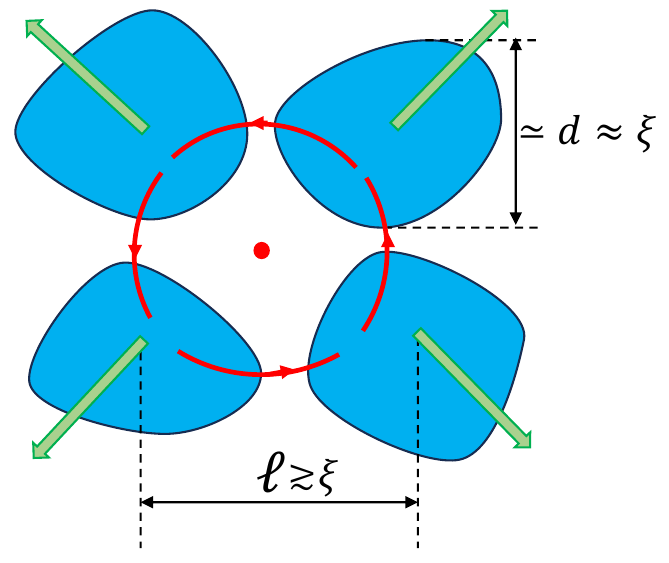}
		\vspace{-0.3cm}
		\caption{A 2D XY vortex in a granular superconductor. The superconducting granules are shown in blue. Green arrows stand for the local order parameter at each granule, $\Psi=|\Psi|e^{{\mathrm i}\varphi}$. Red arrows indicate the phase change between neighboring granules, $\varphi\rightarrow\varphi+\pi/2$, so that upon full circulation around the singularity point shown by a small red circle, the phase changes by $2\pi$.  Such vortices have only gauge structure and no dissipative core with suppressed (neither completely nor even partially) order parameter.}
		\label{Fig1}
	\end{figure}

	So what are the remaining configurations? Let us consider the random JJA model of the granular superconductor. Each island is characterized by an independent phase associated with its local condensate. We can thus have configurations in which these phases realize a non-trivial $2\pi$ circulation over the neighboring islands around a plaquette of the array, as shown in Fig.\,\ref{Fig1}. This corresponds to a vortex without core; such a vortex can be viewed as a point excitation associated with a site of the dual array, exactly as a vortex in the XY model, see\,\cite{minnhagen} for a review. The only relevant scale characterizing this vortex is the typical intergrain distance $\ell$, which plays the role of the ultraviolet cutoff in the model. 

\begin{figure}[t!]
		\includegraphics[width=7cm]{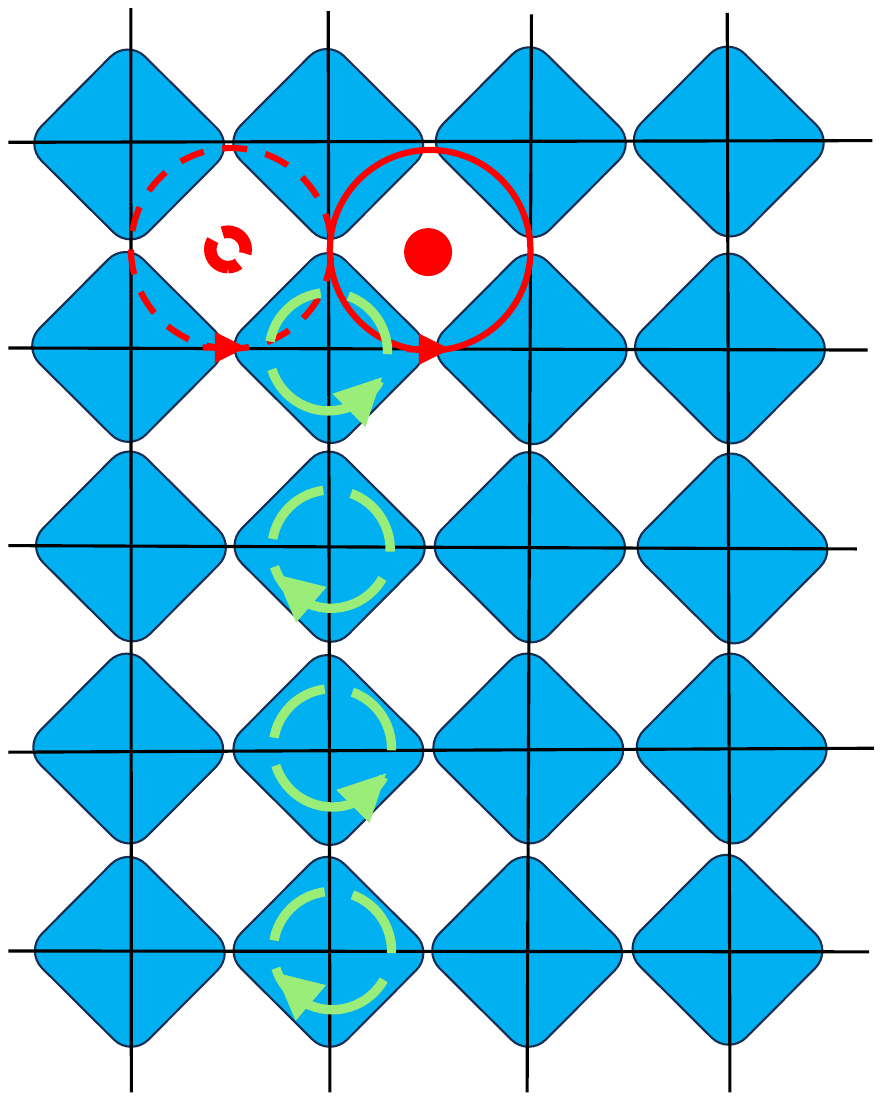}
		\vspace{-0.3cm}
		\caption{A 2D version of a quantum phase slip, corresponding to a line of usual phase slips of alternating sign, ending on a given superconducting grain. This event corresponds to the tunnelling of one vortex between two sites of the dual lattice adjacent to this last grain.} 
		\label{Fig2}
	\end{figure}

	An essential characteristics of these core-less, point-like XY vortices is their extreme quantum mobility. This crucial aspect is usually neglected when focusing only on the classical statistical mechanics of the XY model\,\cite{ischia}. At the quantum level, the phases on the superconducting granules can flip by a $2\pi$ angle, exactly as in the case of quantum phase slips on quantum wires or Josephson junction chains\,\cite{golubev, arutyunov}. Notably important are certain coherent flips over several granules. Let us consider, in particular, a quantum event in which all phases on a line ending in a particular granule simultaneously flip by $\pm 2\pi$, with alternating signs on neighboring granules, as shown in Fig.\,\ref{Fig2}. No phase degrees of freedom, neither small fluctuations (like spin waves), nor vortices, are involved in such an event, except at the end point, where it corresponds to the tunnelling of a single vortex from one plaquette to the neighboring one, see Fig.\,\ref{Fig2}. This is the 2D equivalent of the usual 1D quantum phase slips\,\cite{golubev, arutyunov}. As a consequence of the 2D quantum phase slips, XY vortices can tunnel over the dual array, exactly as Cooper pairs can tunnel between granules of the original array, as was originally pointed out in\,\cite{dst}. We stress again that no dissipative core is involved in this process, the vortices that are tunnelling are the point excitations of the XY model. One might think that the phase flip on one island might excite quasi-particles out of the local condensate there. However, this is not the case; at low temperatures, dissipation due to quasi-particle excitation is negligible, since the typical tunnelling frequency is much smaller than the excitation gap for quasi-particles. The dissipation picture of\,\cite{ioffe} is also not valid, since it is fully predicated on a vortex liquid in the Bose metal state and the latest experiments have now confirmed the original picture\,\cite{dst} in which vortices are mostly frozen in this state, and are not in a liquid phase\,\cite{marcus}. 
	
	Of course, vortex tunnelling creates quantum dissipation for charges in the orthogonal direction and vice versa. But these effects are represented by effective Coulomb potentials (giving rise to related effective `electric fields') for both charges and vortices and can thus be incorporated dynamically into a gauge theory as Gauss law constraints, giving rise exactly to the frozen topological state\,\cite{dst,nodisorder}. The upshot is that quantum effects cannot be neglected for the low-temperature behaviour of granular superconductors, in 2D as well as in 1D. What is normally misinterpreted as the effect of disorder is simply the frustration caused by the orthogonal tunnelling of both charges and vortices. 

	\section{Mutual statistics of charges and vortices. Topological gauge theory}~~\\
		\vspace{-0.8cm}
	
	\noindent
	As we have shown, the Ginzburg-Landau theory cannot describe global superconductivity in these emergent electronic granular systems. Now we demonstrate that the effective field theory that replaces the standard GL approach is a topological gauge theory. As derived above, the normal state of the global superconductor, where the tunneling charge  does not percolate through the sample, involves mobile charges and vortices that can both tunnel locally through the system. The resulting charge and vortex dynamics can be described by a conserved integer-valued three-current $Q_{\mu}$ for point charges, measured in units of $2e$, and either an integer-valued conserved three-current of point vortices $M_{\mu}$ in 2D or an integer-valued conserved antisymmetric four-rank tensor of 1D extended vortices $M_{\mu \nu}$ in 3D, measured in units of $2\pi/2e$. The crucial point is that out of condensate charges and vortices are subject to mutual statistics interactions\,\cite{wilczek}, which stem from nothing else than from the Aharonov-Bohm\,\cite{aharonovbohm} and Aharonov-Casher\,\cite{aharonovcasher} phases acquired by the wave functions of charges and vortices when moving in the presence of the other type of excitation. Exactly as quantum phase slips, these are also quantum effects which cannot be forgotten when dealing with the physics of granular superconducting films at low temperatures. In particular, Cooper pairs and vortices are mutual fermions. 
	
	In order to encode the mutual statistics interactions in a local field theory one must introduce two fictitious gauge fields\,\cite{wilczek}, a vector gauge field $a_{\mu}$ coupling to the charge current and either a pseudovector gauge field $b_{\mu}$ coupling to the vortex current, in 2D, or a pseudotensor antisymmetric gauge field $b_{\mu \nu}$ coupling to the vortex pseudotensor current in 3D\,\cite{semenoff}. The actions for these gauge fields are
	\begin{eqnarray}
		&&S_{2D} = \int d^3x \ {1\over 2\pi} a_{\mu} \epsilon^{\mu \alpha \nu} \partial_{\alpha} b_{\nu} -a_{\mu} Q^{\mu} - b_{\mu} M^{\mu} \ ,
		\nonumber \\
		&&S_{3D} = \int d^4x \ {1\over 4\pi} a_{\mu} \epsilon^{\mu \alpha \nu \rho} \partial_{\alpha} b_{\nu \rho} - a_{\mu} Q^{\mu} -{1\over 2} b_{\mu \nu} M^{\mu \nu} \ ,
		\label{actions}
	\end{eqnarray}
where we use Einstein notation in which a sum over equal Greek space-time indices is implied. 

The emergent fields are invariant under the gauge transformations
\begin{eqnarray}
a_{\mu} &&\to a_{\mu} + \partial_{\mu} \chi \ ,
\nonumber \\
b_{\mu} &&\to b_{\mu} + \partial_{\mu} \lambda \qquad \qquad \qquad \qquad \ 2D \ ,
\nonumber \\
b_{\mu \nu} &&\to b_{\mu \nu} + \partial_{\mu} \lambda_{\nu} - \partial_{\nu} \lambda_{\mu}  \qquad \qquad 3D \ ,
\label{gt}
\end{eqnarray}
the last line representing a so-called gauge transformation of the second kind for a tensor gauge field coupling to 1D vortices. Since $a_{\mu}$ is a vector field and $b_{\mu}$ (or respectively $b_{\mu \nu}$) is a pseudovector  (respectively a pseudotensor) field, the model is invariant under both parity and time-reversal transformations. 
	
Once it is established that gauge fields are necessary for an adequate description of vortices in granular media, the correct effective field theory can be constructed, as usual, by adding all possible gauge invariant terms in a derivative expansion \cite{topsc, planar, typeIII}. This is the same procedure as the one that has been used by Ginzburg and Landau in the case of a scalar field. 
	
The infrared-dominant terms in (\ref{actions}) are the famous Chern-Simons\,\cite{jackiw2} and BF\,\cite{bowick} topological terms of first-order in derivatives. Note that the equations of motion imply that the gauge fields themselves encode the charge and vortex currents,
\begin{eqnarray}
Q^{\mu} &&= {1\over 2\pi} \epsilon^{\mu \alpha \nu} \partial_{\alpha} b_{\nu} \ ,
\nonumber \\
M^{\mu} &&= {1\over 2\pi} \epsilon^{\mu \alpha \nu} \partial_{\alpha} a_{\nu} \qquad \qquad \ \ 2D \ ,
\nonumber \\
M^{\mu \nu} &&= {1\over 4\pi} \epsilon^{\mu \nu \alpha \beta} \partial_{\alpha} a_{\beta }\qquad \qquad 3D \ .
\label{classcurr}
\end{eqnarray}

The next-order terms in the effective action must be gauge-invariant and of second-order in derivatives. For a usual vector gauge field this is the Maxwell term
\begin{eqnarray}
S_{\rm M} &&= -{1\over 4f^2} \int d^3 x \ f_{\mu \nu} f^{\mu \nu} \ ,
\nonumber \\
S_{\rm M} &&= -{1\over 4g^2} \int d^3 x \ g_{\mu \nu} g^{\mu \nu} \qquad \qquad 2D \ ,
\label{maxwell}
\end{eqnarray}
where $f^2$ and $g^2$, both of canonical dimension mass in 2D and dimensionless in 3D are the effective electric and magnetic coupling strengths in the material and non-relativistic effects can be incorporated by a light velocity $v<1$. Here 
\begin{equation}
f_{\mu \nu} = \partial_{\mu } a_{\nu} -\partial_{\nu} a_{\mu} \ ,
\label{fields}
\end{equation}
is the field strength tensor associated with the gauge field $a_{\mu}$ and a corresponding expression holds for $g_{\mu \nu}$ in terms of $b_{\mu}$. In 3D, the field strength tensor is a three-tensor, the so-called Kalb-Ramond tensor, 
\begin{equation}
h_{\mu \nu \alpha} = \partial_{\mu} b_{\nu \alpha} + \partial_{\nu} b_{\alpha \mu}+ \partial_{\alpha} b_{\mu \nu}  \qquad 3D \ ,
\label{kb}
\end{equation}
and the gauge invariant, second-order in derivatives term is given by
\begin{equation}
S_{\rm KR} = {1\over 12 \Lambda^2} \int d^4x \ h_{\mu \nu \alpha}h^{\mu \nu \alpha} \qquad \ 3D \ ,
\label{3Dac}
\end{equation}
where $\Lambda$ is a coupling with canonical dimension mass. 

Normally, the Maxwell and Kalb-Ramond actions alone describe massless gauge fields. The infrared-dominant topological terms in (\ref{actions}), however, gives them a gauge-invariant mass\,\cite{jackiw2, bowick}. Their presence has the consequence that in the normal state the interactions between charges and vortices are screened Yukawa interactions.
The total effective actions for the charge-vortex system, including the coupling of the electric current in (\ref{classcurr}) to the real electromagnetic gauge potential $A_{\mu}$ are thus
\begin{eqnarray}
S_{\rm 2D} &&= \int d^3x \ {-1\over 4f^2} f_{\mu \nu}f^{\mu \nu} + {1\over 2\pi} a_{\mu} \epsilon^{\mu \alpha \nu} \partial_{\alpha} b_{\nu} +{-1\over 4g^2} g_{\mu \nu} g^{\mu \nu} 
\nonumber \\
&&- a_{\mu} Q^{\mu} - b_{\mu} M^{\mu} + {1\over 2\pi} A_{\mu} \epsilon^{\mu \alpha \nu} \partial_{\alpha} b_{\nu} \ ,
\nonumber \\
S_{3D} &&= \int d^4x \ {-1\over 4f^2} f_{\mu \nu} + {1\over 4\pi} a_{\mu} \epsilon^{\mu \alpha \nu \rho} \partial_{\alpha} b_{\nu \rho} +{1\over 12 \Lambda^2} h_{\mu \alpha \nu}h^{\mu \alpha \nu} 
\nonumber \\
&&- a_{\mu} Q^{\mu} -{1\over 2} b_{\mu \nu} M^{\mu \nu} + {1\over 4\pi} A_{\mu} \epsilon^{\mu \alpha \rho \nu} \partial_{\alpha} b_{\rho \nu} \ .
\label{totalactions}
\end{eqnarray}

The global superconducting state is realized when the charges condense, which, in this case, means that electric tunnelling currents can percolate through the effective JJA. Then one can substitute the integer-valued point current $Q_{\mu}$ by a continuous field\,\cite{book}. Remembering that charges are conserved, we can thus introduce the field $q_{\mu}$ defined by $Q^{\mu} = \epsilon^{\mu \alpha \nu} \partial_{\alpha} q_{\nu}$. Defining further a shifted field ${q^\prime}_{\mu} = q_{\mu} -b_{\mu}$ and integrating over it, we see that, in this phase, the original emergent gauge field $a_{\mu}$ decouples completely. We are thus left with the topological gauge which represent the equivalent of the GL representation for type-III superconductors,
\begin{eqnarray} 
S^{\rm SC}_{\rm 2D} &&=  \int d^3 x \  {1\over 2\pi} b_{\mu} \left( F^{\mu} - 2\pi M^{\mu} \right) 
-{1\over 4g^2} g_{\mu \nu} g^{\mu \nu}\nonumber\\
&& - {1\over f^2} F_{\mu \nu}F^{\mu \nu} \ ,
\nonumber \\
S^{\rm SC}_{\rm 3D} &&=  \int d^4 x \ {1\over 4\pi} b_{\mu \nu} \left( \tilde F^{\mu \nu} - 2\pi M^{\mu \nu} \right) 
+ {1\over 12 \Lambda^2} h_{\mu \nu \rho} h^{\mu \nu \rho} 
\nonumber \\
&& - {1\over f^2} F_{\mu \nu}F^{\mu \nu} \ ,
\label{nonrelac3}
\end{eqnarray}
where $F_{\mu} = \epsilon_{\mu \alpha \nu} \partial_{\alpha} A_{\nu}$ is the dual field strength in 2D and $\tilde F_{\mu \nu} = \epsilon^{\mu \nu \rho \lambda} \partial_{\rho} A_{\lambda}$ the corresponding dual field strength in 3D. This shows that, in a type-III superconductor, the gauge gap $1/\lambda$ is opened by a topological mass generation mechanism instead of the usual Anderson-Higgs mechanism. The penetration depth $\lambda$ plays the role of a gauge-invariant infrared cutoff, typically of the order of the system lateral size or larger. 
In 2D, the gauge gap $1/\lambda_{\rm 2D} = O(fg)$, while in 3D $1/\lambda_{\rm 3D} = O(f\Lambda)$. Note that there is no scalar field involved and, correspondingly there is no coherence length. Accordingly, these are non-interacting field theories, quadratic in the fields. The role of the ultraviolet cutoff is taken by the typical inter-grain distance $\ell$ and can be incorporated explicitly by formulating these gauge theories on a lattice modelling the grain structure. While emergent granularity has been known as a paradigm of planar superconductivity since the very early days\,\cite{goldman1}, this granularity is not confined to 2D systems. Recently, bulk superconductors with an emergent granularity have been experimentally detected\,\cite{parra}. Moreover, emergent granularity is a characteristics of the high-$T_{\mathrm c}$ superconductors, especially in the underdoped regime\,\cite{barisic}.

\section{Logarithmic confinement and generalized BKT mechanism}~~\\
		\vspace{-0.8cm}
	
	\noindent
The effective actions (\ref{nonrelac3}) show a crucial feature of the vortex physics of type-III superconductors. For distances $l < \lambda$, the emergent gauge fields $b_{\mu}$ in 2D or $b_{\mu \nu}$ in 3D imply a Coulomb interaction for the vortex currents $M_{\mu}$ in (2+1)-dimensional space-time and $M_{\mu \nu}$ in (3+1)-dimensional space time. This is tantamount to logarithmic confinement of vortices, both in 2D and in 3D\,\cite{dst}. Therefore, the destruction of global superconductivity in type-III superconductors, associated with the suppression of percolating electric tunnelling currents on the effective JJA, is caused by the liberation of logarithmically confined vortices. In 2D, this is nothing else than the famed BKT physics\,\cite{ber, kt}. Our results show, however, that this mechanism is not confined to 2D but, rather, the same physics applies also in 3D. The space dimension is not important, what is relevant for this type of superconductivity is the granularity of the material. Of course, the logarithmic confinement of vortices can be derived also directly in the neutral model\,\cite{typeIII}, exactly as in 2D superfluids\,\cite{minnhagen}. 

The vortex Coulomb interaction has another important consequence. This long-range interaction is the cause of the divergence of a single vortex energy, in 2D, or energy per unit length, in 3D, with the logarithm of the sample size\,\cite{ber, kt, typeIII}. If we apply a uniform magnetic field to the sample, however, this plays for vortices the same role as a neutralizing background for a Coulomb gas\,\cite{minnhagen}. The Debye-H\"uckel mechanism will thus screen the Coulomb interaction. Assuming Maxwell-Boltzmann statistics for the vortices, the ensuing Debye screening length approaches the ultraviolet cutoff $\ell$ as $T\to 0$. As a consequence, the self-energy $E_{\rm vortex} \propto {\rm log} (\lambda / \xi)$ of Abrikosov vortices\,\cite{tinkham} is replaced by $E_{\rm vortex} \propto {\rm log} (\lambda_{\rm Debye} / \ell)$, which vanishes for $T \to 0$ so that the lower critical field $H_{\rm c1}$ of type-III superconductors also vanishes in this limit,
	\begin{equation}
		\lim_{{\rm T} \to 0} H_{\rm c1} = 0 \ .
		\label{zerolower}
	\end{equation}\\
The vortices, however, will still form an Abrikosov-like lattice\,\cite{ameur}. 
	
As we have pointed out above, type-III superconductivity is destroyed by vortex deconfinement and not by the Cooper pairs breaking up, as in usual type-I and type-II superconductors. In 2D this leads to the famous BKT scaling
	\begin{equation}
		R_{\square} \propto {\rm e}^{-\sqrt {b \over |T-T_{\rm BKT}|}}
		\label{bkt}
	\end{equation}
	of the sheet resistance at the transition temperature $T_{\rm BKT}$, where $b$ is a constant having the temperature dimensionality\,\cite{ber,kt}. This is the typical behavior of the 2D XY model \cite{minnhagen} and, therefore, one might be lead to guess that, in 3D, the corresponding behavior is that of the 3D XY model. This is wrong though: the present model does not involve only phase degrees of freedom, as in the XY model, but also the dynamics of charges. The normal state is not a traditional metal but a Bose metal in which freed vortices and charges form a frozen topological state \cite{dst, nodisorder}. Correspondingly, the transition is in the universality class of confining strings \cite{quevedo}. In 2D, this happens to coincide with the XY class. In 3D, however, the vortices are 1D-extended objects. This leads to a modified Vogel-Fulcher-Tamman (VFT) scaling of the resistance at the transition\,\cite{vft},
	\begin{equation}
		R \propto {\rm e}^{-{b^{\prime} \over |T-T_{\rm BKT}|} }\ .
		\label{vft}
	\end{equation}
	This behavior has recently been clearly detected in bulk samples of NbN and NbTiN\,\cite{typeIII}. 
	
	When charges exit the condensate at the transition, the mutual statistical interaction with the liberated vortices generates a gauge-invariant topological mass for these liberated vortices\,%latter\\
	\cite{dst}. The original Coulomb interaction of vortices is thus screened and the resulting energy of vortices is overwhelmed by their entropy in any dimension, causing their proliferation. In 3D, type-III vortices are thus magnetic confining strings, with an action induced by massive antisymmetric gauge fields\,\cite{quevedo}. Correspondingly, the transition destroying type-III superconductivity is the Hagedorn transition, see, e.g.,\,\cite{hagedorn}, for these strings. The vortex free energy per unit length $F\left( \beta \right) $ at temperatures $T=1/\beta$ above this transition can be computed in the large $D$ approximation\,\cite{high1, high2} as 
	\begin{equation}
		F^2\left(\beta \right) \propto - {D-2\over \beta^2} \ .
		\label{free}
	\end{equation} 
	The negative sign indicates that entropy fluctuations indeed overwhelm the string tension and favour long strings. As pointed out in\,\cite{polchinski}, this result can also be interpreted that more and more degrees of freedom are excited as we raise the temperature. 
	
	A very interesting property of these type-III vortices in 3D arises when the antisymmetric tensor action above the transition contains a topological term \cite{quevedo},
	\begin{equation}
		S_{\theta} = {\theta \over 32 \pi^2} \int d^4x \ b_{\mu \nu} \epsilon^{\mu \nu \alpha \beta} b_{\alpha \beta}  \ ,
		\label{theta}
	\end{equation}
	where $\theta$ is an angular variable. In this case, the induced vortex action acquires a topological term 
	\begin{equation}
		S_{\rm vortex}^{\rm top} = -i {\pi t \over (4\pi^2 / f^2)^2 + t^2} \ ,
		\label{selfinter}
	\end{equation}
	where $f^2$ is the effective electric coupling in the material, $t= \theta /\pi$ and $\nu$ is the self-intersection number of the vortex world-sheet in 4D Euclidean space. In the strong coupling limit $f^2 \to \infty$ and for $\theta =\pi$, the weight of each self-intersection in the partition function is (-1): the vortices become fermionic \cite{polya}. 
	
	\section{Conclusion}~~\\
\vspace{-1.1cm}

	\noindent
	To conclude, we would like to stress that the essence of type-III superconductivity is that a single massless superfluid mode can always be described in terms of a generalized gauge field in any dimension. In granular media, type-III vortices are structureless point particles that couple directly to this gauge field, and, as always in gauge theories, a Coulomb interaction is generated by the Gauss law constraint. Furthermore, the vortex coupling automatically generates a photon gap as a gauge invariant topological mass.  A single generalized gauge theory describing, in particular, type-III vortices accommodates all superconducting phenomena and replaces the Ginzburg-Landau theory with no need for a scalar field and the associated coherence length. 
	
	\noindent
	\textit{Acknowledgements}–– The work of VMV is supported by Terra Quantum AG.
	%\section*{References}~~\\
	%	\noindent \textbf{References}
	\vspace{-0.6cm}
%	\bibliographystyle{plain}
	%\bibliography{main}
	
	\bigskip

%	\section*{References}~~\\
		
																\end{document}